\documentclass[useAMS,usenatbib,referee]{biom}
\usepackage{rotating}
\usepackage{amsmath, amssymb} 
\usepackage{algorithm} 
\usepackage{algpseudocode} 
\usepackage{xcolor}
\usepackage{mathrsfs}
\def\bSig\mathbf{\Sigma}

\title[TiVAC Correlation]{A Time-Varying and Covariate-Dependent Correlation Model for Multivariate Longitudinal Studies}

\author{
Qingzhi Liu$^{1}$, 
Gen Li$^{1}$,
Anastasia K. Yocum$^{2}$,
Melvin McInnis$^{2}$,
Brian D. Athey$^{2,3}$,\\
\textbf{Veerabhadran Baladandayuthapani}$^{1,*}$\email{veerab@umich.edu} \\
$^{1}$Department of Biostatistics, University of Michigan, Ann Arbor, Michigan, U.S.A. \\
$^{2}$Department of Psychiatry, University of Michigan, Ann Arbor, Michigan, U.S.A. \\
$^{3}$Department of Computational Medicine and Bioinformatics, University of Michigan, \\ Ann Arbor, Michigan, U.S.A.
}

\begin{document}





\volume{}
\pubyear{}
\artmonth{}


\doi{}


\label{firstpage}


\begin{abstract}
In multivariate longitudinal studies, associations between outcomes often exhibit time-varying and individual level heterogeneity, motivating the modeling of correlations as an explicit function of time and covariates. However, most existing methods for correlation analysis fail to simultaneously capture the time-varying and covariate-dependent effects. We propose a Time-Varying and Covariate-Dependent (TiVAC) correlation model that jointly allows covariate effects on correlation to change flexibly and smoothly across time. TiVAC employs a bivariate Gaussian model where the covariate-dependent correlations are modeled semiparametrically using penalized splines. We develop a penalized maximum likelihood-based Newton–Raphson algorithm, and inference on time-varying effects is provided through simultaneous confidence bands. Simulation studies show that TiVAC consistently outperforms existing methods in accurately estimating correlations across a wide range of settings, including binary and continuous covariates, sparse to dense observation schedules, and across diverse correlation trajectory patterns. We apply TiVAC to a psychiatric case study of 291 bipolar I patients, modeling the time-varying correlation between depression and anxiety scores as a function of their clinical variables. Our analyses reveal significant heterogeneity associated with gender and nervous-system medication use, which varies with age, revealing the complex dynamic relationship between depression and anxiety in bipolar disorders.
\end{abstract}

%



\maketitle


%

\section{INTRODUCTION}
\label{s:intro}
\begin{figure}[h]
\centering
\includegraphics[width=\textwidth]{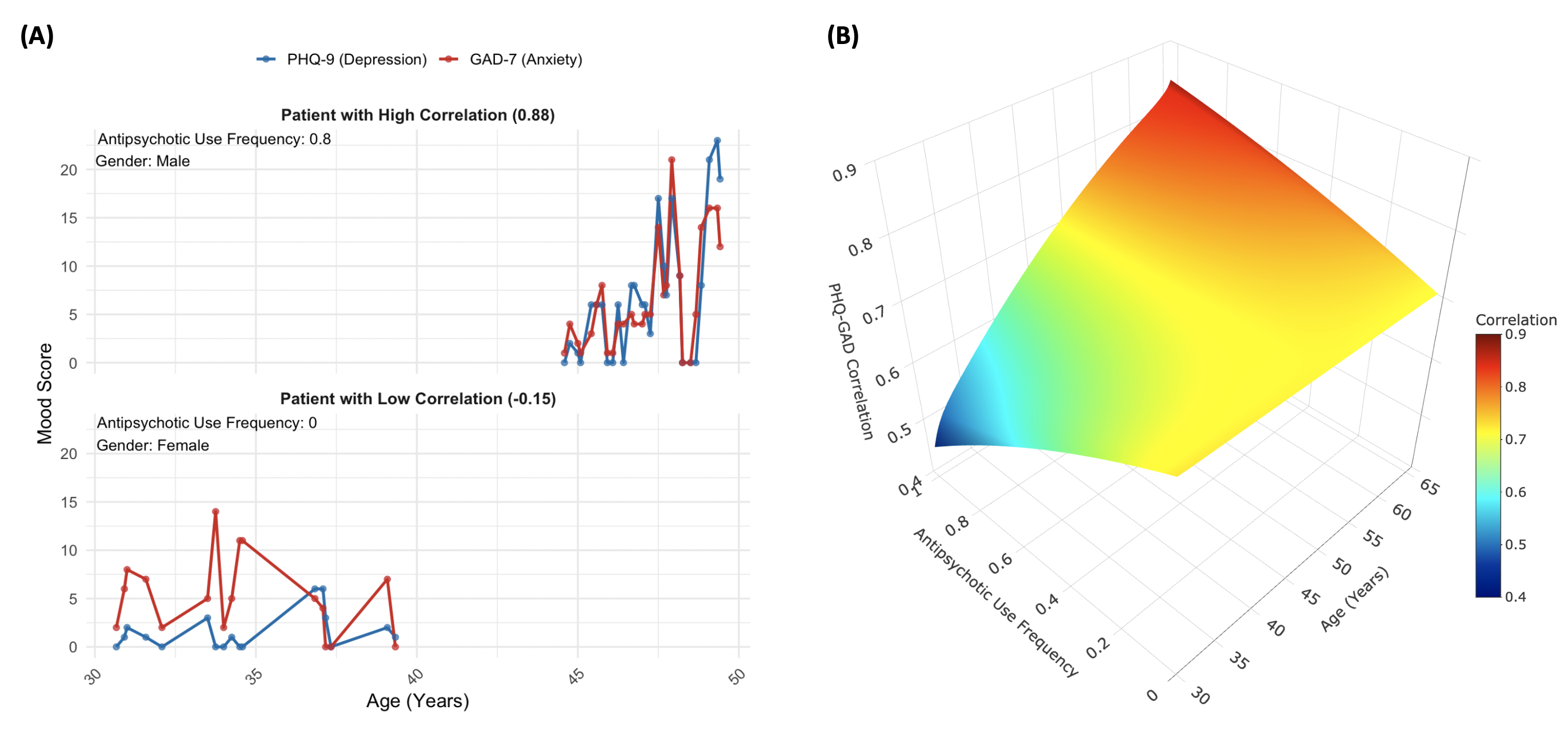}
\caption[PHQ-9 and GAD-7 trajectories and correlation heterogeneity]{\textbf{PHQ-9 and GAD-7 trajectories and correlation heterogeneity.}
\textbf{(A)} Irregular longitudinal trajectories of PHQ-9 (depression; blue) and GAD-7 (anxiety; red) for two representative BP-I patients. The concurrent, within-patient Pearson correlation is high for the top patient ($\rho=0.88$) and low for the bottom patient ($\rho=-0.15$).
\textbf{(B)} Three-dimensional correlation surface showing how the PHQ–GAD correlation (z-axis) varies with age (x-axis) and antipsychotic-use frequency (y-axis). Color encodes the correlation value through rainbow scale.}
\label{TiVAC_Surface}
\end{figure}

Multivariate longitudinal data often reveal that associations between outcomes are neither constant over time nor uniform across individuals. As an example, in psychiatric research, depression and anxiety are commonly regarded as highly comorbid conditions\citep{Kalin_2020}. Interestingly, in our application to bipolar I disorder (BP-I) patients, the concurrent correlation between PHQ-9 (depression score; \cite{Kroenke_2001}) and GAD-7 (anxiety score; \cite{Spitzer_2006}) was heterogeneous across patients, with some exhibiting very high correlations and others low or near zero (as shown in Figure~\ref{TiVAC_Surface}A). To further characterize this heterogeneity, we found that the correlation also varied with patient-specific covariates such as age and antipsychotic use frequency, as illustrated by the three-dimensional correlation surface in Figure~\ref{TiVAC_Surface}B. These findings underscore the importance of accounting for both time-varying and covariate-dependent heterogeneity in correlation, that could manifest in other contexts as well. This point is also highlighted by the ubiquitous Simpson’s paradox: subgroup-level correlations can vanish or even reverse relative to a fixed population-level correlation \citep{Blyth_1972}. This motivates the statistical question of modeling correlations of outcomes ($\rho$) as an explicit function of time ($t$) and covariates ($\mathbf{X}$), as $\rho(t,\mathbf{X})$ in multivariate longitudinal studies.

In the context of covariate-dependent correlations (i.e. $\rho(\mathbf{X})$), \citet{Bartlett_1993} first considered correlation as a function of a discrete third variable. However, this method requires large subgroup-specific sample sizes and is limited to a single discrete covariate \citep{Dufera_2023}. Subsequent extensions have allowed correlation to depend on both continuous and discrete covariates within a bivariate Gaussian framework, with parameters estimated via restricted maximum likelihood (REML) \citep{Wilding_2011} and second-order generalized estimating equations (GEE2) \citep{Ho_2010}. More recently, \citet{Tu_2022} introduced the covariate-dependent pairwise correlation model with association size (CoCoA) and compared the performance of REML and GEE2 estimators with maximum likelihood estimators (MLE). Another broad class of methods focuses on modeling covariate-dependent covariance or correlation for more than two outcomes through parameterizations or transformations that ensure the covariance or correlation matrix remains positive definite \citep{Pourahmadi_1999, Chiu_1996, Hoff_Niu_2012, Ni_2019, Archakov_2021}.

Despite these advancements, the aforementioned models do not explicitly address time-varying correlations. Estimating covariate-dependent correlations separately at each time point can substantially reduce statistical power, particularly when data are irregularly observed. Although one might consider treating time as a single covariate \citep{Wilding_2011, Tu_2022}, this strategy often imposes a linearity assumption on the time-varying effects, which may obscure potentially dynamic and nonlinear patterns in the correlation structure. For instance, in our motivating application, Web Figure 4 displays the monthly Pearson correlation between depression and anxiety severity across different ages. When examining the full cohort or stratifying by gender, the estimated correlation trajectories reveal significant noise that masks the underlying true dynamic pattern, thereby motivating the use of a flexible nonlinear modeling approach. Although several statistical approaches--including DCC-GARCH \citep{Engle_2002}, DSEM \citep{Asparouhov_2017}, and idPAC \citep{Kundu_2021}--have been proposed to capture time-varying relationships among variables ($\rho(t)$), none directly or jointly model how covariates affect correlations over time and provide statistical inference on these time-varying effects, i.e. $\rho(t,\mathbf{X})$.

To this end, we propose a Time-Varying and Covariate-Dependent (TiVAC) correlation model, which jointly models time-varying and covariate-dependent correlations. TiVAC focuses on concurrent correlations between pairs of outcomes over time and facilitates statistical inference on how covariate effects on correlation might change with time. Under a bivariate Gaussian model, we apply a Fisher transformation to the correlation parameter and formulate the transformed correlation as a linear function of covariates with coefficients that vary over time. To capture the dynamic nature of these coefficients, we employ functional data analysis methods using spline-based constructions, which offer two main benefits: flexible modeling of covariate and time effects, and borrowing strength across adjacent time points. For estimation, we develop a computationally efficient strategy incorporating penalization to ensure spline smoothness and interpretability, and maximize the penalized likelihood with a Newton–Raphson algorithm. To test nonlinear time-varying and covariate effects, we construct simultaneous confidence bands to control for multiple testing. In simulations, TiVAC consistently outperforms several competing state-of-the-art approaches in estimating $\rho(t,\mathbf{X})$ across various time-dependent correlation functions and covariate types, under both correctly specified and misspecified models. We further applied TiVAC to mood trajectory data from 291 BP-I patients enrolled in the Prechter Longitudinal Study of Bipolar Disorder (PLS-BD) \citep{mcinnis2018cohort,Yocum_2023}, examining how patient characteristics and clinical interventions were associated with the correlation between PHQ-9 and GAD-7 scores across patients' age. Our results revealed that the overall correlation between depression and anxiety increased slightly with age. Notably, males exhibited higher correlations than females at younger ages. Additionally, while higher usage of nervous system-related medications and antipsychotics was linked to lower correlations in early adulthood, this relationship reversed at older ages; in contrast, mood stabilizer use was consistently associated with lower correlations.

The remainder of the paper is organized as follows. Section~\ref{sec:Proposed Model} details the TiVAC model formulation. Sections~\ref{s:Estimation_Spline_Representation} and \ref{sec:Spline Coefficient Estimation} describe the spline representation and estimation procedure of coefficients, and Section~\ref{sec:Simultaneous Confidence Band Estimation} outlines the construction of simultaneous confidence bands for inference. Simulation results for correctly specified data are presented in Section~\ref{sec: Simulations with Deterministic Correlation Functions}, while Section~\ref{sec: Impact of Noise in Correlation Functions} investigates model performance under incorrect specifications. Section~\ref{s:Application} demonstrates the application of TiVAC in the BP-I patient mood trajectory study. Section~\ref{s:discuss} concludes with a discussion of our findings and future research directions.

\section{TiVAC MODEL}
\label{sec:Proposed Model}

We first outline the data structure and the TiVAC model framework. The observed data consist of two main components: a pair of concurrent and irregular longitudinal outcomes and a set of \(p\) covariates. Specifically, let \(n\) denote the number of individuals. For each individual \(i\), let  
\begin{equation*}
\mathbf{Y}_i = \bigl[\{\,Y_{i1}(t_{i1}),\,Y_{i2}(t_{i1})\,\},\,\dots,\,\{\,Y_{i1}(t_{im_i}),\,Y_{i2}(t_{im_i})\,\}\bigr]
\end{equation*}
represent the vector of observed outcomes, where \(Y_{i1}(t_{ij})\) and \(Y_{i2}(t_{ij})\) are the first and second outcomes for individual \(i\) at time \(t_{ij}\), \(m_i\) is the number of observation times for individual \(i\), and \(t_{ij}\) denotes the \(j\)-th observation time for individual \(i\). Let \(\mathbf{Y} = (\mathbf{Y}_1^T, \dots, \mathbf{Y}_n^T)\) represent the collection of observed outcomes for all individuals. Additionally, let \(\mathbf{Y}_i(t_{ij}) = \{\,Y_{i1}(t_{ij}),\, Y_{i2}(t_{ij})\,\}^T\) represent the bivariate outcome pair for individual \(i\) at time \(t_{ij}\). The covariate information is represented by a design matrix $\mathbf{X}$ of dimensions $n \times p$, where $n > p$ (with $p$ low-dimensional), and $\boldsymbol{X}_i$ is the $p$-dimensional covariate vector for individual $i$.

For each individual \(i\) at given time \(t_{ij}\), we assume the concurrently observed outcomes follow a bivariate normal distribution with a covariate-dependent and time-varying correlation:
\begin{equation} 
\small{
\label{eq:bivariate_normal} \begin{bmatrix} Y_{i1}(t_{ij}) \\ Y_{i2}(t_{ij}) \end{bmatrix} \sim \mathcal{N}\bigl\{\mathbf{0}, \boldsymbol{\Sigma}(t_{ij}, \boldsymbol{X}_i)\bigr\}, \quad\text{where}\quad \boldsymbol{\Sigma}(t_{ij}, \boldsymbol{X}_i) = \begin{bmatrix} \sigma_1^2 & \rho(t_{ij}, \boldsymbol{X}_i)\,\sigma_1\,\sigma_2 \\ \rho(t_{ij}, \boldsymbol{X}_i)\,\sigma_1\,\sigma_2 & \sigma_2^2 \end{bmatrix}. 
}
\end{equation} 
Here, \(\rho(t_{ij}, \boldsymbol{X}_i)\) referred to as the TiVAC correlation, depends explicitly on both the observation time \(t_{ij}\) and the covariate vector \(\boldsymbol{X}_i\). We emphasize the dynamic nature of concurrent correlation by allowing \(\rho(t_{ij}, \boldsymbol{X}_i)\) to vary continuously over time. 

Since \(\rho\) is constrained within the interval \([-1,1]\), the Fisher transformation is employed to map it onto the real line as, \[ \eta(t_{ij}, \boldsymbol{X}_i) = \ln\!\left\{\frac{1 + \rho(t_{ij}, \boldsymbol{X}_i)}{1 - \rho(t_{ij}, \boldsymbol{X}_i)}\right\}. \] The transformed correlation \(\eta(t_{ij}, \boldsymbol{X}_i)\) is then modeled via a varying-coefficient linear form as,
\begin{equation} 
\label{eq:fisher_transformation} \eta(t_{ij}, \boldsymbol{X}_i) = \boldsymbol{X}_i^T\,\boldsymbol{\beta}(t_{ij}), \end{equation} 
where \(\boldsymbol{\beta}(t_{ij})\) is a coefficient vector function of time. Of particular interest, the \(k\)-th element of \(\boldsymbol{\beta}(t_{ij})\), denoted as \(\beta_k(t_{ij})\), represents the time-varying effect of the \(k\)-th covariate on the correlation. This specification allows the concurrent correlation to
depend jointly on time and covariates through interpretable and nonlinear coefficient trajectories, and can be construed as a generalization of CoCoA model \citep{Tu_2022}.

Building on the specification above, we impose the following additional assumptions to ensure identifiability and interpretability of TiVAC model: (1) each outcome pair has zero mean, which can be achieved, for example, through a flexible de-meaning approach such as the one proposed by \citet{Patil_2022}; (2) the marginal variances are constant over time ($\sigma_1^2,\sigma_2^2$); since
$\mathrm{Cov}\{Y_{i1}(t),Y_{i2}(t)\}=\rho(t,\boldsymbol{X}_i)\sigma_1(t)\sigma_2(t)$, time-varying marginal variances would confound marginal scale with dependence, complicating inference on $\rho(t,\boldsymbol{X}_i)$, especially under irregular designs (cf.~\citet{Kundu_2021}); and (3) there is no autocorrelation across distinct time points, thereby focusing on dynamic heterogeneity in the concurrent correlation while balancing model complexity and computational efficiency.

\section{SPLINE CONSTRUCTION AND ESTIMATION}
\label{s:Estimation}
To obtain a flexible and smooth estimator \(\hat{\beta}_k(t)\), we represent the coefficient functions using B-splines \citep{DE_BOOR_2001}, with known basis functions and spline coefficients to be estimated, as described in Section~\ref{s:Estimation_Spline_Representation}. Because direct maximum-likelihood estimation of spline coefficients is sensitive to the number and placement of knots, we incorporate a smoothing penalty on the spline coefficients, ensuring stability while retaining the ability to capture significant local changes (Section~\ref{sec:Spline Coefficient Estimation}). 
Finally, to address multiple comparisons arising from evaluating covariate effects simultaneously across numerous time points, we develop a bootstrap-based inference framework to construct simultaneous confidence bands, ensuring that the entire trajectory of covariate effects lies within the bands with a specified confidence level (Section~\ref{sec:Simultaneous Confidence Band Estimation}).

\subsection{Spline representation of nonlinear covariate effects}
\label{s:Estimation_Spline_Representation}
Direct estimation of \(\boldsymbol{\beta}(t_{ij})\) is not feasible, as it is infinite-dimensional. To address this, we approximate it using a finite-dimensional B-spline representation \citep{DE_BOOR_2001}. B-splines are piecewise polynomial functions, with the order of the spline determining the degree of the polynomial; a common choice is cubic B-splines (order 4). The number and placement of knots control the flexibility of the spline; more knots allow for greater flexibility, while fewer knots lead to smoother estimates. Specifically, let \(\boldsymbol{b}(t_{ij})\) denote a \(q\)-dimensional vector of B-spline basis functions evaluated at time \(t_{ij}\), where \(q\) depends on the number of knots and the spline order. For the \(i\)-th individual, the B-spline basis functions evaluated at all observed time points are organized into a spline basis matrix, $\mathbf{B}_i=\big\{\boldsymbol{b}(t_{i1})\ \cdots\ \boldsymbol{b}(t_{im_i})\big\}^{\!\top}\in\mathbb{R}^{m_i\times q}$.

\textit{Incorporating B-splines into TiVAC.} For each covariate \(k\), the corresponding time-varying coefficient \(\beta_k(t_{ij})\) in \(\boldsymbol{\beta}(t_{ij})\) can then be expressed as:
\begin{equation}
\label{eq:beta_spline_representation}
\beta_k(t_{ij}) = \boldsymbol{b}(t_{ij})^T \boldsymbol{\theta}_k,
\end{equation}
where \(\boldsymbol{\theta}_k\) is a (lower) \(q\)-dimensional vector of spline coefficients. Substituting (\ref{eq:beta_spline_representation}) into (\ref{eq:fisher_transformation}), the only parameters to estimate are the spline coefficients, \(\boldsymbol{\theta}_k\), for \(k = 1, \dots, p\). To simplify the representation, we combine the spline bases and covariates into a single unified structure. Specifically, for the \(i\)-th individual at time \(t_{ij}\), we define a new covariate vector \(\mathbf{A}_{i,j} = \boldsymbol{X}_i^T \otimes \boldsymbol{b}(t_{ij})\) with length \(q \cdot p\), where \(\otimes\) denotes the Kronecker product. Let \(\boldsymbol{\theta} = (\boldsymbol{\theta}_1^T, \boldsymbol{\theta}_2^T, \dots, \boldsymbol{\theta}_p^T)^T\), a vector of length \(q \cdot p\), represent all spline coefficients across the \(p\) covariates. The Fisher-transformed correlation for individual \(i\) at time \(t_{ij}\) can then be expressed as:
\begin{equation*}
\eta(t_{ij}, \boldsymbol{X}_i) = \mathbf{A}_{i,j}^T \boldsymbol{\theta}.
\end{equation*}
After \(\boldsymbol{\theta}\) is estimated, it can be substituted into Equation~\ref{eq:beta_spline_representation} to obtain the interpretable coefficient functions \(\beta_k(t_{ij})\).

\subsection{Spline coefficient estimation}
\label{sec:Spline Coefficient Estimation}

\textit{Smoothness penalty on spline coefficients.} To estimate the spline coefficients \(\boldsymbol{\theta}\), we maximize a penalized log-likelihood function that incorporates a smoothness penalty on the B-spline coefficients. The penalty term ensures stability in the estimation process and prevents overfitting. Rather than using the traditional integral of squared derivatives, we adopt a difference penalty on adjacent spline coefficients, as proposed by \citet{Eilers_1996}. This discrete penalty simplifies computation while providing a close approximation to the smoothness measure achieved by integral-based methods. Specifically, we employ the second-order difference penalty, defined as:
\begin{equation*}
P(\boldsymbol{\theta}_k) = \frac{1}{2} \lambda_k \boldsymbol{\theta}_k^T \mathbf{D}_2^T \mathbf{D}_2 \boldsymbol{\theta}_k,
\end{equation*}
where \(\mathbf{D}_2\) represents the second-order difference operator matrix, and \(\lambda_k > 0\) is the smoothing parameter. Specifically, \(\mathbf{D}_2 \boldsymbol{\theta}_k\) yields a vector whose \(j\)-th element is \(\theta_{k,j} - 2\theta_{k,j+1} + \theta_{k,j+2}\). Thus, this penalty encourages adjacent spline coefficients to remain close, penalizing curvature and promoting smoothness in the estimated coefficient functions. Higher-order difference penalties can be easily extended within this framework. As suggested by \citet{Eilers_1996}, using a relatively large number of knots is advisable. This strategy eliminates the need to balance low versus high numbers of knots, which in unpenalized methods can yield drastically different smoothness levels. However, we caution against using an excessively dense grid of knots, as this can inflate the number of parameters and lead to insufficient data between adjacent knots, potentially complicating model fitting.

\textit{Estimation of spline coefficients.} Incorporating the smoothness penalty, the penalized log-likelihood function for the spline coefficients is given by:
\begin{equation}
\label{eq:penalized log-likelihood}
l_{\text{penalized}}(\boldsymbol{\theta}; \mathbf{Y}, \mathbf{A}) = \log L(\boldsymbol{\theta}; \mathbf{Y}, \mathbf{A}) - \frac{1}{2} \sum_{k=1}^p \lambda_k \boldsymbol{\theta}_k^T \mathbf{D}_2^T \mathbf{D}_2 \boldsymbol{\theta}_k.
\end{equation}
Here, \(\log L(\boldsymbol{\theta}; \mathbf{Y}, \mathbf{A})\) represents the log-likelihood function for all individuals, incorporating their \(m_i\) bivariate outcomes:
\begin{equation*}
\log L(\boldsymbol{\theta}; \mathbf{Y}, \mathbf{A}) = -\frac{1}{2} \sum_{i=1}^n \sum_{j=1}^{m_i} \{  \log(2\pi) + \log |\boldsymbol{\Sigma}(t_{ij}, \boldsymbol{X}_i)| + \mathbf{Y}_i(t_{ij})^T \boldsymbol{\Sigma}(t_{ij}, \boldsymbol{X}_i)^{-1} \mathbf{Y}_i(t_{ij}) \}.
\end{equation*}
To estimate the spline coefficients \(\boldsymbol{\theta}\), we maximize the penalized log-likelihood function defined in (\ref{eq:penalized log-likelihood}) with respect to \(\boldsymbol{\theta}\), given certain penalized parameters \(\lambda_k\), \(k = 1, \dots, p\):
\begin{equation*}
\hat{\boldsymbol{\theta}} = \underset{\boldsymbol{\theta}}{\arg \max} \ l_{\text{penalized}}(\boldsymbol{\theta}; \mathbf{Y}, \mathbf{A}).
\end{equation*}
We employ the Newton–Raphson algorithm to iteratively update \(\boldsymbol{\theta}\). In addition, the smoothing parameters $\{\lambda_k\}_{k=1}^p$ are optimized via cross-validation by maximizing the held-out log-likelihood \(l(\hat{\boldsymbol{\theta}}; \mathbf{Y}, \mathbf{A})\). Further algorithmic details are providedin Web Appendix A.1.

\textit{Estimation of the TiVAC correlation.}  After obtaining the optimal smoothing parameters, the model is fit using all available samples to estimate the optimal \(\hat{\boldsymbol{\theta}}\). Let \(\mathbf{B}\) be the spline basis matrix evaluated at all time points from \(1\) to \(\max_{i,j}(t_{ij})\), having dimensions \(\max_{i,j}(t_{ij}) \times q\), where \(\max_{i,j}(t_{ij})\) denotes the maximum observed time across all individuals. The estimated coefficient trajectory \(\hat{\beta}_k(t)\), for \(t=1,\ldots,\max_{i,j}(t_{ij})\), is given by
\begin{equation*}
\hat{\beta}_k(t) = \mathbf{B}\,\hat{\boldsymbol{\theta}}_k,
\end{equation*}
which measures the \(k\)-th covariate's time-varying effect on the correlation. The correlation trajectory for covariate \(\boldsymbol{X}_i\) can then be expressed as:
\begin{equation*}
\hat{\rho}(t, \boldsymbol{X}_i) = \tanh\!\left\{\sum_k X_{ik} \hat{\beta}_k(t)\right\},
\end{equation*}
where \(X_{ik}\) is the \(k^{th}\) covariate for individual \(i\).

\subsection{Simultaneous confidence band estimation and inference}
\label{sec:Simultaneous Confidence Band Estimation}

After estimating \(\beta_k(t)\), it is important to conduct inference to determine whether the covariate effect on the correlation at certain time points is significant. Instead of using pointwise confidence bands, which only guarantee confidence at individual time points, we use simultaneous confidence bands that ensure the entire trajectory of a function lies within the band with a specified confidence level. Mathematically, a \(100(1-\alpha)\%\) simultaneous confidence band satisfies:
\begin{equation*}
P \left\{ L(t) \leq \beta_k(t) \leq U(t), \, \forall \, t \in T \right\} \geq 1-\alpha \text{ for all } t \in T.
\end{equation*}
This approach is crucial for addressing multiple comparison issues and provides broader coverage compared to pointwise confidence bands.

We construct simultaneous confidence bands for \(\beta_k(t)\) using a bootstrap procedure that resamples individuals (with replacement), preserving the observed time points within each individual. Specifically, our procedure consists of two nested bootstrap loops, each serving a distinct purpose. The outer bootstrap loop quantifies the variability of the estimator \(\hat{\beta}_k(t)\) by repeatedly drawing bootstrap samples \(\mathbf{Y}^{(b)}\) (\(b = 1,\dots,B\)) and re-estimating the spline coefficients, yielding bootstrap estimates \(\hat{\beta}_k^{(b)}(t)\). To avoid underestimating the variability of the outer bootstrap estimates, we use an inner bootstrap loop to provide robust estimates of their standard deviations. Specifically, in each outer iteration \(b\), we perform \(M\) additional bootstrap resamples \(\mathbf{Y}^{(b,m)}\) (\(m = 1,\dots,M\)) from the outer sample \(\mathbf{Y}^{(b)}\), computing inner-loop estimates \(\hat{\beta}_k^{(b,m)}(t)\). From these inner bootstrap samples, we compute the sample standard deviation \(\tilde{s}_{k,t}\) for each outer iteration. Next, we compute the t-statistic for each outer iteration \(b\) following \citet{Ruppert_2010}:
$
T_k^{(b)} = \max_t \left|\frac{\hat{\beta}_k^{(b)}(t) - \hat{\beta}_k(t)}{\tilde{s}_{k,t}}\right|.
$
The critical value \(T_k^{\text{crit}}\) is then obtained as the empirical \((1-\alpha)\)-quantile of the collection of t-statistics \(\{T_k^{(b)}\}_{b=1}^{B}\). As noted by \citet{Ruppert_2010}, when the sample size is large, \(T_k^{\text{crit}}\) can be approximated with high accuracy for simultaneous confidence band estimation. Finally, the simultaneous confidence band is constructed as:
$
\hat{\beta}_k(t) \pm T_k^{\text{crit}} \cdot s_{k,t},
$
where \(s_{k,t}\) is the overall bootstrap standard deviation of \(\hat{\beta}_k^{(b)}(t)\) across all outer bootstrap iterations \(b = 1, \dots, B\). A detailed summary of the entire procedure is provided in Web Appendix A.2.

\section{SIMULATION STUDIES}  
\label{s:Simulation Section}  

We evaluate the estimation accuracy of the TiVAC correlation under four generative models, characterized by whether the univariate covariate is binary or continuous and whether the underlying correlation structure follows a deterministic function or is influenced by random noise. For penalty parameter selection, we employ 10-fold cross-validation to determine the optimal value.

We compare the TiVAC correlation to three alternative methods. The first is the ``Empirical" correlation trajectory, calculated by computing the Pearson correlation between the two outcomes at each observed time point within each covariate group. The resulting correlations are then smoothed over time using LOESS regression, with candidate span values ranging from 0.1 to 1 in increments of 0.1. The optimal span is selected via 5-fold cross-validation. The second method is covariance regression (CovReg) \citep{Hoff_Niu_2012}, where the covariate-dependent covariance matrix of the two outcomes is calculated at each time point, transformed into a correlation matrix, and subsequently smoothed using the same LOESS approach as in the Empirical method. The third method is CoCoA \citep{Tu_2022}, implemented with three different estimation techniques: MLE, REML \citep{Wilding_2011}, and GEE2 \citep{Ho_2010}. The resulting estimates are also smoothed using the same LOESS approach as in the Empirical method. Following the recommendation of \citet{Tu_2022} to use CoCoA-REML, we focus primarily on its performance in the main text; results for CoCoA-MLE and CoCoA-GEE2 are provided in Web Appendix B.

For the simulation settings, we fix the sample size at \(n = 300\) and set \(\max_{i,j}(t_{ij}) = 500\). Each individual \(i\) has \(m_i\) observed time points drawn under one of four conditions: (1) a fixed 40 time points (\(\mathscr{T}_{High}\)); (2) a random number of time points between 3 and 40 (\(\mathscr{T}_{Moderate}\)), mimicking the observed PLS-BD data; (3) a random number of time points between 3 and 10 (\(\mathscr{T}_{Low}\)); and (4) an alternative setting in which the number of individuals and time points matches those of the observed PLS-BD data, with \(\max_{i,j}(t_{ij}) = 421\) (see Section~\ref{s:Application}), denoted \(\mathscr{T}_{PLS-BD}\). The outcomes \(Y_{i1}(t_{ij})\) and \(Y_{i2}(t_{ij})\) follow a bivariate normal distribution (Equation~\ref{eq:bivariate_normal}) with variances \(\sigma_1^2 = 1\) and \(\sigma_2^2 = 4\).

To generate the time- and covariate-dependent correlation \(\rho(t_{ij}, X_i)\), we apply the Fisher transformation in two separate data-generating models: a deterministic model (correctly specified case),
\begin{equation*}
\eta(t_{ij}, X_i) \;=\; \beta_0(t_{ij}) \;+\; \beta_1(t_{ij})\,X_i,
\end{equation*}
and a noisy model (misspecified case),
\begin{equation*}
\eta(t_{ij}, X_i) \;=\; \beta_0(t_{ij}) \;+\; \beta_1(t_{ij})\,X_i \;+\; \epsilon(t_{ij}),
\end{equation*}
where \(\epsilon(t_{ij}) \sim N\bigl(0,\,\sigma_{\text{noise}}\bigr)\), and $\sigma_{\text{noise}}\in\{0.1,0.2,0.3,0.4,0.5\}$. 

The four simulation scenarios considered in Sections~\ref{sec: Simulations with Deterministic Correlation Functions} and \ref{sec: Impact of Noise in Correlation Functions} are:
\begin{itemize}
\item Scenario 1: deterministic model with a binary covariate $X_i\in\{0,1\}$;
\item Scenario 2: deterministic model with a continuous covariate $X_i\in[0,1]$;
\item Scenario 3: noisy model with a binary covariate $X_i\in\{0,1\}$;
\item Scenario 4: noisy model with a continuous covariate $X_i\in[0,1]$.
\end{itemize}

The coefficient functions \(\beta_k(t)\) \((k = 0,1)\) are generated using one of three trajectory patterns: linear, seasonal, or logistic. The true correlation curves, \(\tanh\{\beta_0(t) + \beta_1(t)\}\), are shown as dashed lines in Figure~\ref{f:Scenario1_Figure}.

For model evaluation, we compute the root mean squared error (RMSE) by comparing each method’s estimated correlation to the true correlation curve over 50 replications. In the binary covariate scenario, RMSE is averaged across time points for each covariate value (0 or 1), reflecting the deviation from the true correlation trajectory in each group. In the continuous covariate scenario, RMSE is computed by averaging over both time points and a grid of 100 covariate values ranging from 0 to 1.

\subsection{Simulations with deterministic correlation functions}  
\label{sec: Simulations with Deterministic Correlation Functions}

In Scenario 1, the underlying correlation function is deterministic, and the covariate is binary (\(X = 0\) for half the sample and \(X = 1\) for the other half). Panel A of Table~\ref{t:Simulation_Table} presents the RMSE and standard deviations (SD) results for TiVAC, Empirical, CovReg, and CoCoA-REML across different coefficient function types and time-point settings. Overall, TiVAC achieves the highest estimation accuracy under all settings, with its advantage becoming more pronounced when fewer time points are observed. Under \(\mathscr{T}_{High}\), the RMSE differences between TiVAC and the other methods are modest (up to 0.02). By contrast, in the \(\mathscr{T}_{Low}\) setting, Empirical, CovReg, and CoCoA-REML exceed TiVAC’s RMSE by more than 0.1 in most cases. TiVAC also exhibits smaller SD than the competing methods, reflecting its stability. TiVAC’s outperformance could stem from its capacity to borrow correlation information across time points via penalized splines. Under the \(\mathscr{T}_{Low}\) setting, CovReg performs worse than Empirical, and CoCoA-REML exhibits even poorer performance. We suspect that CovReg and CoCoA-REML, as model-based approaches, are particularly sensitive to data sparsity. In practice, when only a few observations are available at certain time points, the iterative optimization in model-based methods may fail to converge, leading to missing estimates of correlation.

\begin{table}[H]
\caption{\textbf{Simulation results for Scenarios 1 and 2.} Simulation results comparing the performance of TiVAC and competing methods under different coefficient function types and number of observed time points per individual. Panel A (Scenario 1) presents root mean squared error (RMSE) with standard deviation (SD) in parentheses when the covariate is binary; the time-varying correlation is \(\rho_0(t)\) when \(X = 0\) and \(\rho_1(t)\) when \(X = 1\). Panel B (Scenario 2) presents RMSE (SD) when the covariate is continuous. }
\label{t:Simulation_Table}
\small
\begin{center}

Panel A: Scenario 1 (binary covariate)\\[0.25em]
\begin{tabular}{lllcccc}
\hline
Coefficient & \# of Time & Correlation & \multicolumn{4}{c}{RMSE (SD)} \\
\cline{4-7} \rule{0pt}{2.5ex}
Functions & Points & Functions & TiVAC & Empirical & CovReg & CoCoA-REML \\
\hline
Linear   & \(\mathscr{T}_{Low}\)      & \( \rho_0(t) \) & 0.02 (0.01) & 0.16 (0.04) & 0.20 (0.10) & 0.36 (0.03) \\ 
         &                            & \( \rho_1(t) \) & 0.02 (0.01) & 0.15 (0.04) & 0.20 (0.09) & 0.34 (0.04) \\ 
         & \(\mathscr{T}_{Moderate}\) & \( \rho_0(t) \) & 0.01 (0.01) & 0.06 (0.02) & 0.04 (0.01) & 0.06 (0.02) \\ 
         &                            & \( \rho_1(t) \) & 0.01 (0.01) & 0.06 (0.02) & 0.04 (0.02) & 0.07 (0.02) \\ 
         & \(\mathscr{T}_{High}\)     & \( \rho_0(t) \) & 0.01 (0.01) & 0.03 (0.01) & 0.02 (0.01) & 0.03 (0.01) \\ 
         &                            & \( \rho_1(t) \) & 0.01 (0.01) & 0.03 (0.01) & 0.02 (0.01) & 0.02 (0.01) \\ 
         & \(\mathscr{T}_{PLS-BD}\)     & \( \rho_0(t) \) & 0.01 (0.01) & 0.04 (0.02) & 0.04 (0.02) & 0.09 (0.02) \\ 
         &                            & \( \rho_1(t) \) & 0.02 (0.01) & 0.06 (0.02) & 0.05 (0.02) & 0.10 (0.02) \\ 
\hline
Seasonal & \(\mathscr{T}_{Low}\)      & \( \rho_0(t) \) & 0.06 (0.02) & 0.12 (0.04) & 0.17 (0.07) & 0.08 (0.03) \\ 
         &                            & \( \rho_1(t) \) & 0.06 (0.02) & 0.16 (0.03) & 0.19 (0.06) & 0.22 (0.03) \\ 
         & \(\mathscr{T}_{Moderate}\) & \( \rho_0(t) \) & 0.04 (0.01) & 0.06 (0.02) & 0.07 (0.02) & 0.05 (0.02) \\ 
         &                            & \( \rho_1(t) \) & 0.03 (0.01) & 0.07 (0.02) & 0.07 (0.02) & 0.07 (0.01) \\ 
         & \(\mathscr{T}_{High}\)     & \( \rho_0(t) \) & 0.03 (0.01) & 0.04 (0.01) & 0.04 (0.01) & 0.04 (0.01) \\ 
         &                            & \( \rho_1(t) \) & 0.03 (0.01) & 0.04 (0.01) & 0.04 (0.01) & 0.04 (0.01) \\ 
         & \(\mathscr{T}_{PLS-BD}\)     & \( \rho_0(t) \) & 0.03 (0.01) & 0.05 (0.02) & 0.07 (0.02) & 0.05 (0.02) \\ 
         &                            & \( \rho_1(t) \) & 0.03 (0.01) & 0.08 (0.02) & 0.08 (0.02) & 0.08 (0.02) \\ 
\hline
Logistic & \(\mathscr{T}_{Low}\)      & \( \rho_0(t) \) & 0.05 (0.02) & 0.14 (0.04) & 0.19 (0.06) & 0.36 (0.04) \\ 
         &                            & \( \rho_1(t) \) & 0.05 (0.02) & 0.16 (0.05) & 0.22 (0.08) & 0.34 (0.04) \\ 
         & \(\mathscr{T}_{Moderate}\) & \( \rho_0(t) \) & 0.03 (0.01) & 0.06 (0.01) & 0.05 (0.01) & 0.07 (0.02) \\ 
         &                            & \( \rho_1(t) \) & 0.03 (0.01) & 0.06 (0.01) & 0.05 (0.01) & 0.07 (0.02) \\ 
         & \(\mathscr{T}_{High}\)     & \( \rho_0(t) \) & 0.02 (0.01) & 0.03 (0.01) & 0.03 (0.01) & 0.03 (0.01) \\ 
         &                            & \( \rho_1(t) \) & 0.02 (0.01) & 0.03 (0.01) & 0.03 (0.01) & 0.03 (0.01) \\ 
         & \(\mathscr{T}_{PLS-BD}\)     & \( \rho_0(t) \) & 0.03 (0.01) & 0.05 (0.01) & 0.04 (0.02) & 0.10 (0.02) \\ 
         &                            & \( \rho_1(t) \) & 0.03 (0.01) & 0.06 (0.02) & 0.06 (0.02) & 0.09 (0.02) \\ 
\hline
\end{tabular}

\vspace{3em}

Panel B: Scenario 2 (continuous covariate)\\[0.25em]
\begin{tabular}{llccc}
\hline
Coefficient & \# of Time & \multicolumn{3}{c}{RMSE (SD)} \\ 
\cline{3-5} \rule{0pt}{2.5ex}
Functions & Points & TiVAC & CovReg & CoCoA-REML \\ 
\hline
Linear   & \(\mathscr{T}_{Low}\)      & 0.03 (0.01) & 0.21 (0.04) & 0.15 (0.02) \\ 
         & \(\mathscr{T}_{Moderate}\) & 0.02 (0.01) & 0.08 (0.01) & 0.06 (0.01) \\ 
         & \(\mathscr{T}_{High}\)     & 0.01 ($<$0.01) & 0.05 (0.01) & 0.03 (0.01) \\ 
         & \(\mathscr{T}_{PLS-BD}\)     & 0.02 (0.01) & 0.10 (0.01) & 0.09 (0.01) \\ 
\hline
Seasonal & \(\mathscr{T}_{Low}\)      & 0.04 (0.01) & 0.21 (0.05) & 0.12 (0.02) \\ 
         & \(\mathscr{T}_{Moderate}\) & 0.03 (0.01) & 0.09 (0.01) & 0.07 (0.01) \\ 
         & \(\mathscr{T}_{High}\)     & 0.03 ($<$0.01) & 0.05 (0.01) & 0.04 (0.01) \\ 
         & \(\mathscr{T}_{PLS-BD}\)     & 0.03 (0.01) & 0.09 (0.01) & 0.07 (0.01) \\ 
\hline
Logistic & \(\mathscr{T}_{Low}\)      & 0.05 (0.01) & 0.21 (0.04) & 0.16 (0.02) \\ 
         & \(\mathscr{T}_{Moderate}\) & 0.03 (0.01) & 0.09 (0.01) & 0.06 (0.01) \\ 
         & \(\mathscr{T}_{High}\)     & 0.02 ($<$0.01) & 0.05 (0.01) & 0.03 (0.01) \\ 
         & \(\mathscr{T}_{PLS-BD}\)     & 0.03 (0.01) & 0.09 (0.02) & 0.09 (0.01) \\ 
\hline
\end{tabular}

\end{center}
\end{table}

\begin{figure}[H]
\begin{center}
\includegraphics[width=\textwidth]{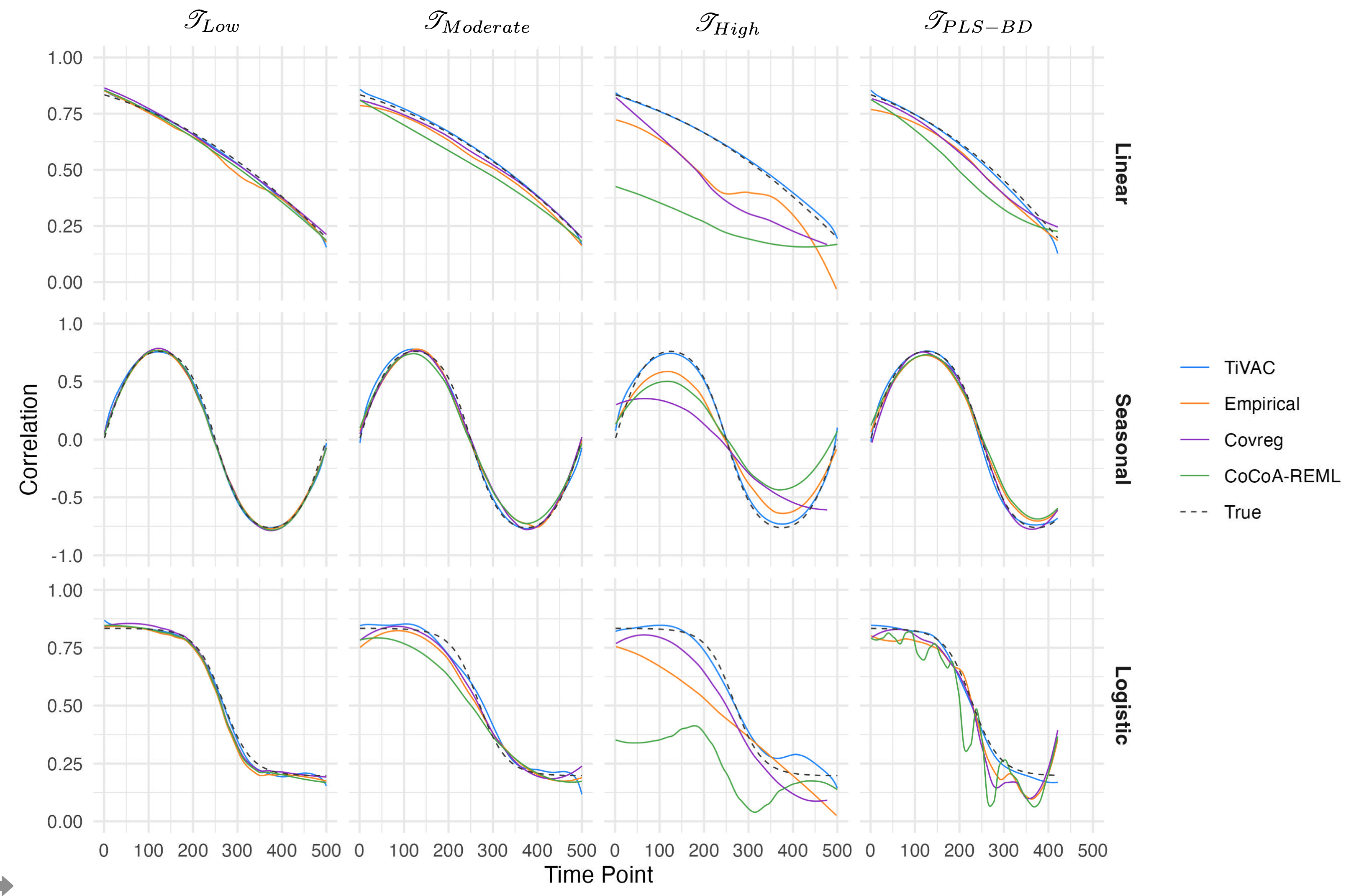}
\end{center}
\caption[Representative curve estimations in Scenario 1.]{\textbf{Representative curve estimations in Scenario 1.} Comparison between the estimated correlation curves from four methods (TiVAC, Empirical, CovReg and CoCoA-REML) and the true correlation curve when covariate \(X = 1\). The figure presents nine independent cases, representing three types of coefficient functions and three settings for the number of time points per individual. Each case shows a representative example from the first run of the simulation out of 50 replications (using a fixed random seed for all cases). The TiVAC, Empirical, CovReg and CoCoA-REML methods are depicted as blue, orange, purple and green curves, respectively, while the true trajectory is represented by a dashed black curve.
\label{f:Scenario1_Figure}}
\end{figure}

Figure~\ref{f:Scenario1_Figure} compares the estimated correlation curves from the four methods to the true curve when \(X = 1\) in Scenario 1. Each subplot corresponds to a representative example from the first of 50 simulation runs (using a fixed random seed). The figure reveals two additional insights. First, TiVAC better captures the tail behavior of the true correlation. For instance, under the \(\mathscr{T}_{PLS-BD}\) logistic trajectory, Empirical, CovReg, and CoCoA-REML exhibit an upward trend on the right tail, whereas the true curve and TiVAC flatten out. Second, TiVAC’s spline-based penalty preserves a level of smoothness that aligns closely with the true trajectory, while pointwise-based estimates can become overly smoothed (e.g., under the \(\mathscr{T}_{Low}\) seasonal trajectory) or fluctuate sharply (e.g., CoCoA-REML under the \(\mathscr{T}_{PLS-BD}\) logistic trajectory). One reason is that the large variability of the pointwise estimates before smoothing makes it difficult for LOESS regression to recover the underlying correlation’s true smoothness.

Furthermore, Web Table 1 presents simulation results for CoCoA-MLE, CoCoA-REML, and CoCoA-GEE2 in Scenario 1. Under the \(\mathscr{T}_{Low}\) setting, CoCoA-GEE2 outperforms the other two estimators, while in other settings their performances are similar. Nevertheless, under the \(\mathscr{T}_{Low}\) setting, its performance remains inferior to that of TiVAC, Empirical, and CovReg.

In Scenario 2, we adopt the same overall simulation design as in Scenario 1 but treat the covariate as continuous, ranging from 0 to 1. Unlike in Scenario 1, we fix each individual’s number of observed time points at 10 to prevent CovReg from failing to converge under the \(\mathscr{T}_{Low}\) setting. Because Empirical cannot accommodate continuous covariates, we focus on TiVAC, CovReg, and CoCoA-REML. Panel B of Table~\ref{t:Simulation_Table} summarizes the RMSE results across different coefficient function types and time-point settings. TiVAC consistently outperforms CovReg and CoCoA-REML. Moreover, TiVAC exhibits similar performance in both Scenario 1 and Scenario 2 when the other settings are comparable, underscoring its robust performance for different covariate types. CoCoA-REML consistently outperforms CovReg, likely because its parametric assumption at each time point aligns with how the data are generated in our simulation, leading to more stable estimates than CovReg in this more complex scenario.

\begin{figure}[H]
\begin{center}
\includegraphics[width=\textwidth]{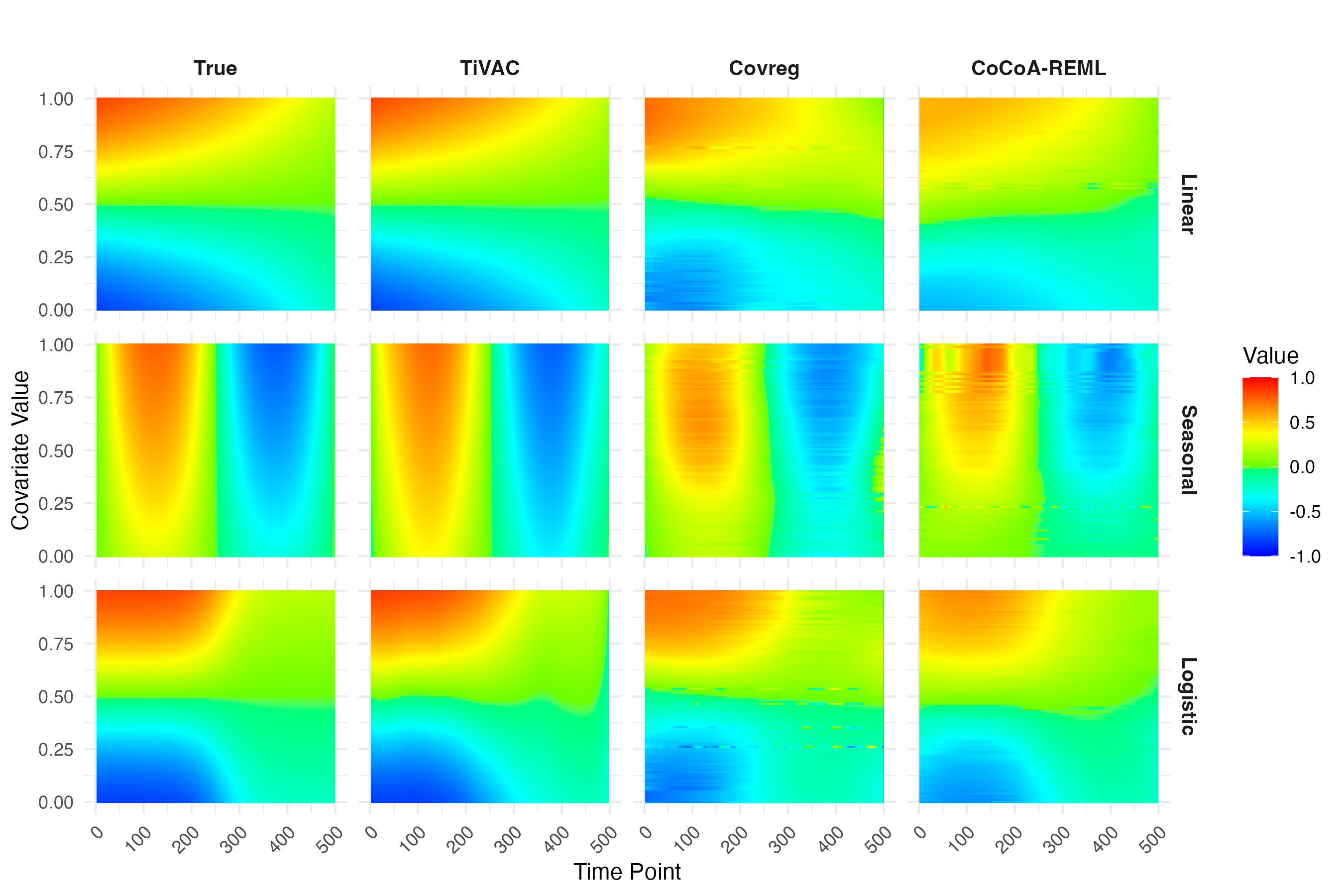}
\end{center}
\caption[Representative heatmap estimations under $\mathscr{T}_{Moderate}$ setting in Scenario 2.]{\textbf{Representative heatmap estimations under $\mathscr{T}_{Moderate}$ setting in Scenario 2.} Comparison between the estimated correlation heatmaps from three methods (TiVAC, CovReg and CoCoA-REML) and the true correlation heatmap under \(\mathscr{T}_{Moderate}\) setting. In each heatmap, correlation values are color-coded, with red indicating higher correlations and blue indicating lower correlations. Each heatmap displays correlation patterns over time for different values of the continuous covariate. The figure presents three independent cases, each representing one of the three types of coefficient functions. Each case shows a representative example from the first run of the simulation (using a fixed random seed for all cases).
\label{f:Scenario2_Figure_Moderate}}
\end{figure}

Figure~\ref{f:Scenario2_Figure_Moderate} shows representative examples from the first run of each case (using a fixed random seed) under the \(\mathscr{T}_{Moderate}\) setting in Scenario 2. The correlation heatmaps illustrate how correlation values (in rainbow colors) depend on both time points (x-axis) and covariate values (y-axis). They reveal that TiVAC’s estimated heatmap is not only much closer to the true pattern but also displays smoother transitions across time and covariates. Additional comparisons of correlation heatmaps under the \(\mathscr{T}_{Low}\), \(\mathscr{T}_{High}\), and \(\mathscr{T}_{PLS-BD}\) settings appear in Web Figures 1-3, respectively. Although CovReg and CoCoA-REML have similar RMSE under \(\mathscr{T}_{Moderate}\) and \(\mathscr{T}_{PLS-BD}\), their heatmaps indicate that these two methods recover the true patterns in \(\mathscr{T}_{PLS-BD}\) more poorly than they do in \(\mathscr{T}_{Moderate}\), illustrating how the irregular nature of the PLS-BD data complicates interpretability of CovReg and CoCoA-REML. Finally, Web Table 2 presents simulation results comparing CoCoA-MLE, CoCoA-REML, and CoCoA-GEE2 in Scenario 2. Unlike Scenario 1, here CoCoA-REML performs better than CoCoA-GEE2 and is comparable to CoCoA-MLE.

\subsection{Impact of noise in correlation functions}  
\label{sec: Impact of Noise in Correlation Functions}

In both Scenario 3 (binary covariate) and Scenario 4 (continuous covariate), we introduce additional randomness into the latent correlation function by letting \(\sigma_{\text{noise}}\) vary from 0.1 to 0.5. We use the \(\mathscr{T}_{Moderate}\) setting and compute the RMSE as in their respective baseline scenarios (i.e., with no random noise in the coefficient function). As summarized in Web Table 3 (for Scenario~3) and Web Table 4 (for Scenario~4), all methods degrade in performance as \(\sigma_{\text{noise}}\) increases, but TiVAC consistently outperforms the alternatives and remains robust, maintaining an average RMSE below 0.1. This robustness likely stems from TiVAC’s spline-based smoothing, which helps mitigate noise-induced fluctuations in the correlation estimates.

Web Table 5 (for Scenario~3) and Web Table 6 (for Scenario~4) compare CoCoA-MLE, CoCoA-REML, and CoCoA-GEE2. In Scenario~3, CoCoA-GEE2 slightly outperforms the other CoCoA estimators for linear and logistic coefficient functions. In Scenario~4, CoCoA-MLE slightly performs better than CoCoA-REML, which in turn outperforms CoCoA-GEE2, for linear and logistic functions; whereas CoCoA-REML slightly outperforms for the seasonal functions.

\section{APPLICATION TO MOOD TRAJECTORIES IN BIPOLAR PATIENTS}
\label{s:Application}
We analyzed data from patients with BP-I enrolled in PLS-BD \citep{mcinnis2018cohort, Yocum_2023}, focusing on the PHQ-9 \citep{Kroenke_2001}, a measure of depression severity, and the GAD-7 \citep{Spitzer_2006}, a measure of anxiety severity, as mood outcomes. Patients' substance use disorder status and medication use were incorporated as covariates. Depression and anxiety are highly prevalent and comorbid symptoms among individuals with bipolar disorder \citep{Kalin_2020}. However, our empirical analysis (Figure~\ref{TiVAC_Surface}A) indicated heterogeneity in their association over time. This pattern motivated use of the TiVAC model to characterize the dynamic correlation between depression and anxiety and to evaluate how patient characteristics and clinical factors modulate this relationship. A better understanding of these patterns could elucidate disease mechanisms and inform personalized treatment strategies \citep{Kim_2024}.

Mood outcomes were collected bimonthly, and medication use records were updated yearly, with substantial between-patient variability in enrollment (baseline) dates and subsequent observation schedules. To ensure a consistent and interpretable timeline, we used each patient’s measured age as the time variable ($t_{ij}$) in the TiVAC model. The dataset was preprocessed (Liu et al. 2025), followed by additional cleaning based on the following inclusion criteria: (1) for each patient, every observed time point must include both PHQ-9 and GAD-7 scores, and (2) the observed time points must correspond to ages between 30 and 65. PHQ-9 and GAD-7 scores were then centered separately for males and females and subsequently transformed using a quantile transformation \citep{Tu_2022} to approximate a Gaussian distribution. These preprocessing steps ensured that the assumptions of the TiVAC model were fairly met. As shown in Web Table 7, the dataset used for modeling included 291 BP-I patients. Of these, 70.8\% were female, 47.4\% had a history of alcohol use disorder, and 28.9\% had cannabis use disorder. Alcohol use disorder and cannabis use disorder status were assessed at enrollment and treated as time-invariant covariates. Medication-related covariates were defined based on (Liu et al. 2025) as follows: the average nervous system-related medications (which include all psychiatric medications) for each patient represented the mean count of those medications across observed time points, while the use frequency of each psychiatric medication reflected the proportion of observed time points where the medication was taken. These covariates were continuous, with this summarization approach necessitated by limitations in the medication data collection process.

We applied the TiVAC model to each covariate separately to measure its average effect on the correlation between PHQ-9 and GAD-7 scores over age in BP-I patients. Although the model can in principle handle multiple covariates, each covariate introduced a set of spline parameters. Because our data were highly irregular (e.g., few observations at certain time points), introducing too many parameters can lead to unstable and unreliable estimates. Age was recorded in months for modeling while visualized in years, with selected knots placed at intervals of 24 months. Without considering covariates, Web Figure 5 shows that the correlation between PHQ-9 and GAD-7 scores slightly increases with age, consistent with the literature \citep{Lenze_2003} suggesting higher comorbidity between PHQ-9 and GAD-7 in older individuals.

\begin{figure}[H]
\begin{center}
\includegraphics[width=0.97\textwidth]{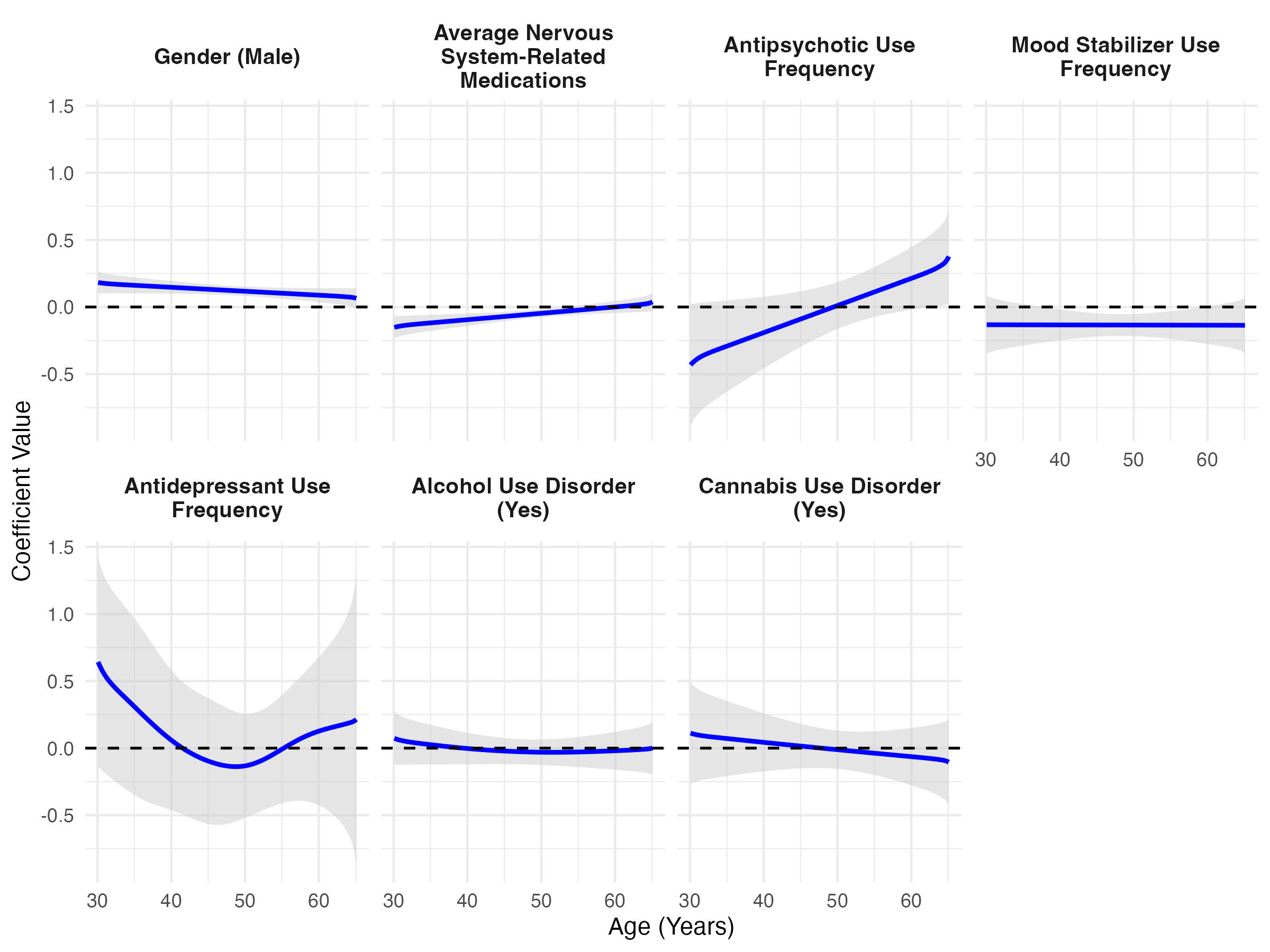}
\end{center}
\caption[Results of TiVAC coefficient functions on PLS-BD data.]{\textbf{Results of TiVAC coefficient functions on PLS-BD data.} TiVAC coefficient functions over age for seven covariates, measuring their effects on the correlation between PHQ-9 and GAD-7 in PLS-BD. For each covariate, the coefficient function $\beta(t)$ is shown in blue, with a gray band representing the 95\% simultaneous confidence band. The dashed black line at zero indicates no effect.
\label{f:TiVAC_Apply_Coefficients}}
\end{figure}

Figure \ref{f:TiVAC_Apply_Coefficients} presents the coefficient functions \(\beta_1(t)\) in blue, representing the effect of each covariate on the correlation as age increases. The 95\% simultaneous confidence bands (shown in gray) indicate the periods where the covariate effect was significant; specifically, when the bands did not overlap with the zero coefficient value (dashed line). For gender, males exhibited significantly higher correlations than females as age increases, though this difference diminished gradually. For the average nervous system-related medications, increased usage was associated with significantly lower correlations until around age 50. Regarding antipsychotic use frequency, greater usage was associated with lower correlations before age 50 and higher correlations after age 50, with the effects being significant at the endpoints of the observed range. For mood stabilizer use frequency, higher usage was significantly associated with lower correlations between ages 40 and 55. Although the estimated coefficient functions for these covariates appeared roughly linear in time, their significance could still vary across the age range; by contrast, assuming a time-invariant effect would imply the significance remains constant. In addition, the effect of antidepressant use frequency exhibited a nonlinear pattern, as shown in its subplot, but there was insufficient evidence to conclude a significant influence over time. Similarly, there was no strong evidence of significant effects for alcohol use disorder or cannabis use disorder. Lastly, the relatively wider confidence bands at the tiles were likely due to less frequent observations in those periods.

\begin{figure}[H]
\begin{center}
\includegraphics[width=\textwidth]{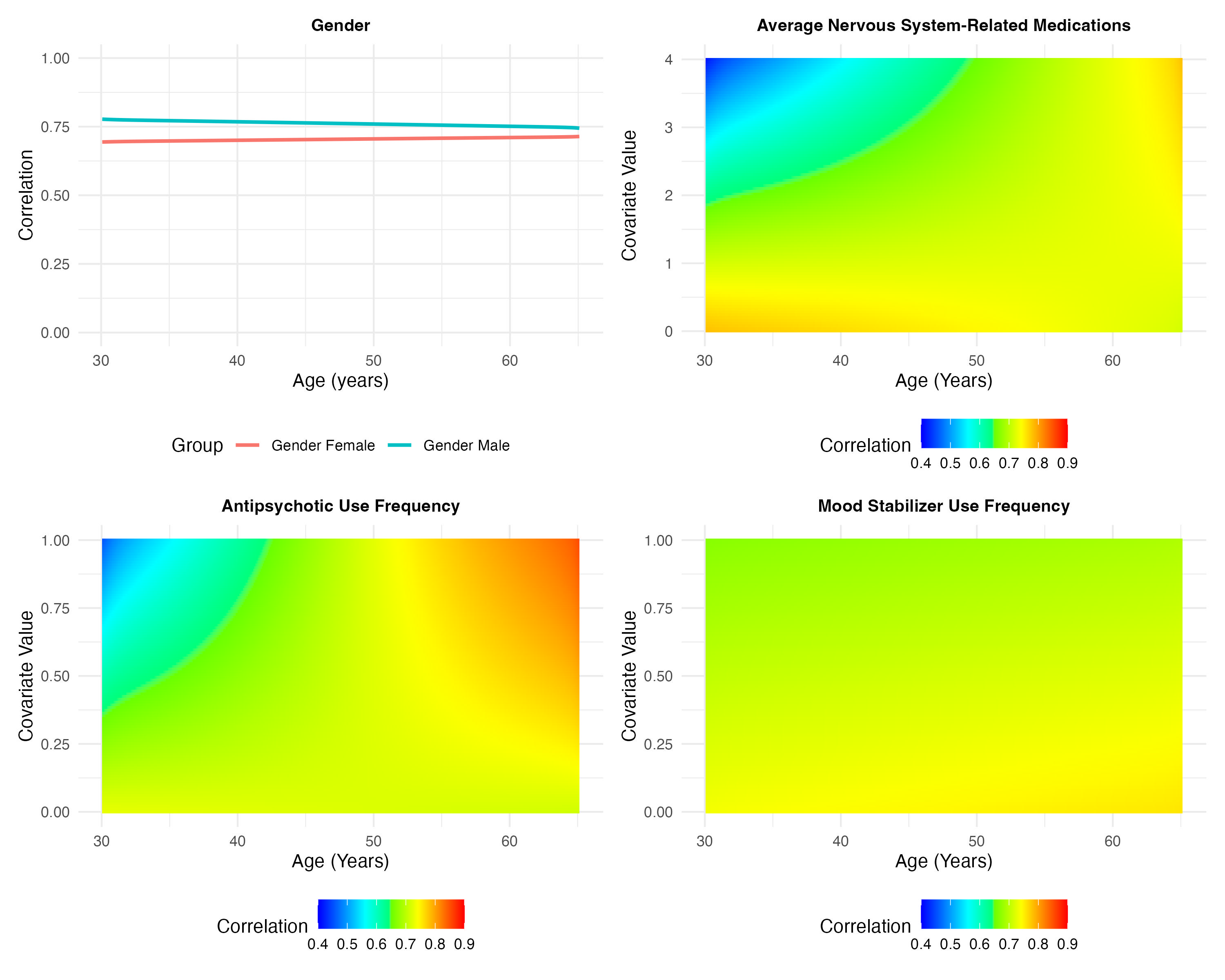}
\end{center}
\caption[Results of TiVAC correlations on PLS-BD data.]{\textbf{Results of TiVAC correlations on PLS-BD data.} Visualization of TiVAC correlations between PHQ-9 and GAD-7 for four covariates with significant or near-significant age periods. For gender (top-left subplot), the TiVAC correlation curve over age is shown in red for females and blue for males. For the other three heatmaps, the corresponding covariates are continuous, with the covariate values on the y-axis, age on the x-axis, and heatmap colors representing the TiVAC correlation values.
\label{f:TiVAC_Apply_Correlations}}
\end{figure}

To further interpret the TiVAC results, we visualized the TiVAC correlations for the four covariates that exhibited significant effects during certain time periods (Figure \ref{f:TiVAC_Apply_Correlations}). Almost all TiVAC correlations across the four subplots exceeded 0.5, with exceptions observed when the average nervous system-related medication usage was near 4 or the antipsychotic use frequency was near 1, particularly around ages 30–35, as indicated in blue. The consistently high TiVAC correlations are consistent with the well-documented comorbidity between anxiety and depressive disorders \citep{Kalin_2020}. Genetically, these disorders share a substantial level of risk factors \citep{Hettema_2008}. At the neural circuit level, both involve alterations in prefrontal-limbic pathways that are critical for mediating emotional regulation processes \citep{Kalin_2020}.

In addition, the TiVAC correlation for males on average was approximately 0.1 higher than that for females at age 30 but gradually diminished as age increases. Previous studies have shown that, in general, women with depression often exhibit greater comorbidity with anxiety disorders than men, likely due to sex differences in neural circuits and molecular mechanisms related to anxiety and depression \citep{Bangasser_2021, Kornstein__2000}. However, this area remains underexplored. Interestingly, within this BP-I cohort, we observed the opposite pattern, underscoring the need for further research into sex-specific mechanisms underlying mood and anxiety comorbidities in BP-I.

For nervous system-related medications, increased usage was associated with lower correlations before age 40 (indicated by a shift from orange to blue) but with higher correlations between ages 60 and 65 (indicated by a shift from green to orange). A similar, though not identical, age-dependent pattern was observed for antipsychotic use frequency, indirectly reflecting differences in drug responses across age groups. One possible explanation involves the brain’s neuroplasticity, which remains relatively high at younger ages \citep{Park_2013}, allowing for greater adaptability. In this context, nervous system-related medications may disrupt natural emotional regulation circuits, thereby weakening correlations between anxiety and depressive symptoms.

In contrast, higher mood stabilizer use frequency was associated with slightly lower correlations (shifting from yellow to green) across all ages. Mood stabilizers are thought to modulate prefrontal-limbic circuits involved in mood regulation \citep{Schloesser_2012}. This stabilizing effect may slightly decouple the neural processes underlying anxiety and depression.

Overall, by applying our TiVAC model to the mood data of BP-I patients, we revealed that the relationship between anxiety and depression depends on gender, medication use, and age, highlighting the complexity of mood dynamics in BP-I patients. While these findings provide data-driven evidence, further validation is necessary to confirm these observations.

\section{DISCUSSION}
\label{s:discuss}

In this work, we develop the Time-Varying and Covariate-Dependent (TiVAC) correlation model, which jointly models time-varying and covariate-dependent correlations. TiVAC accommodates flexible, time-varying covariate effects and time-resolved inference by modeling latent correlation regression coefficients as smooth functions of time. In addition, our bootstrap-based procedure for estimating simultaneous confidence bands mitigates concerns of multiple testing across time points. Simulation studies demonstrate that TiVAC consistently outperforms competing methods in terms of correlation estimation accuracy across various settings. In our application to longitudinal mood data from BP-I patients, TiVAC identified substantial heterogeneity in the association between depression and anxiety across age and patient characteristics. These findings reflect variation in the correlation structure rather than changes in symptom severity or treatment response.

This paper primarily focuses on inference for the TiVAC correlation structure, and an interesting future direction is to jointly model a flexible mean structure together with the TiVAC correlation structure. Currently, TiVAC assumes zero autocorrelation within each outcome across different time points. Extending the bivariate Gaussian framework to a multivariate or Gaussian-process setting would allow for correlations across time-points, but necessitates deeper investigation of the model-building and estimation processes due to increased complexity and computational demands. Another promising direction is to accommodate non-Gaussian outcomes, given that many real-world datasets may exhibit skewness, zero-inflation, or discrete distributions. In this setting, copula-based methods may offer a flexible approach to effectively address these distributional challenges. Finally, although the current framework focuses on two correlated variables, extending TiVAC to higher-dimensional outcomes is a promising avenue for future research; such extensions will require parameterizations that guarantee positive definiteness of the correlation matrix.

When applying TiVAC in practice, several considerations are important for interpreting the results. The framework does not allow for clear temporal separation between exposure and correlation; therefore, it estimates statistical associations rather than predictive or causal effects. Moreover, the magnitude of correlation alone does not directly translate to illness severity, treatment response, or prognosis.

\textit{}

\backmatter


\section*{ACKNOWLEDGMENTS}

The authors thank the Heinz C. Prechter Bipolar Research Program for providing the data. The authors also thank Dr. David Belmonte and Dr. Jie Sun for providing valuable clinical perspectives on the interpretation of the application results.

\section*{FUNDING}
Veerabhadran Baladandayuthapani’s work was supported by the National Institutes of Health grants R01CA244845-01A1 and P30 CA46592 and funds from the University of Michigan Rogel Cancer Center and School of Public Health.

\section*{DATA AVAILABILITY}
The data used in this study were obtained under license from the Heinz C. Prechter Bipolar Research Program and are therefore not publicly available. Data are however available from the authors upon reasonable request and with permission from the Heinz C. Prechter Bipolar Research Program. The code and R package used to reproduce all results presented in this paper are available at https://github.com/bayesrx/TiVAC.

\bibliographystyle{biom} \bibliography{TiVAC}


%


\label{lastpage}

\end{document}